\documentstyle[12pt]{article}
\renewcommand{\theequation}{\arabic{section}.\arabic{equation}}
\newcounter{saveeqn}
\newcommand{\alpheqn}{\setcounter{saveeqn}{\value{equation}}%
\setcounter{equation}{0}\addtocounter{saveeqn}{1}%
\renewcommand{\theequation}{\arabic{section}.\arabic{saveeqn}.
\alph{equation}}}
\newcommand{\reseteqn}{\setcounter{equation}{\value{saveeqn}}%
\renewcommand{\theequation}{\arabic{section}.\arabic{equation}}}
\newcommand{\dslash}{\partial\hskip-6.6pt /\hskip2pt}
\newcommand{\pslash}{p\hskip-6.6pt /\hskip2pt}

\newcommand{\Aslash}{A\hskip-6.7pt /\hskip2pt}
\begin{document}
\begin{titlepage}\hfill HD-THEP-95-38
\vspace*{2cm}
\begin{center}{\Large\bf Optimized $\delta $ - Expansion and
Triviality \\ or Non - Triviality of Field Theories}
\\[5 ex]
{\bf Dieter Gromes}\\[3 ex]Institut f\"ur
Theoretische Physik der Universit\"at Heidelberg\\ Philosophenweg 16,
D-69120 Heidelberg \\ E - mail: D.Gromes@ThPhys.Uni-Heidelberg.DE \\
 \end{center}
\par
\vspace*{3cm}
{\bf Abstract: } We use a very simple version of the optimized
(linear) $\delta $ - expansion by scaling the free part of the
Lagrangian with a variational parameter. This method is well suited to
calculate the renormalized coupling constant in terms of the free one
and the cutoff. One never has to calculate any new Feynman graphs but
simply can modify existing results from the literature. We find that
$\Phi _4^4$ -theory as well as QED are free in the limit where the
cutoff goes to infinity. In contrast to this, the structure of
Yang-Mills theories enforces a special choice of the Lagrangian of the
$\delta $ - expansion. Together with the change in the sign of the
$\beta $ - function, this leads to a different behavior and allows
Yang-Mills theory to become non trivial.
\vfill \centerline{September 1995}
\end{titlepage}
\section{Introduction}
The ``optimized $\delta $ -expansion'' (for the older literature see
Stevenson \cite{stev85} and references therein), also called ``linear
$\delta $ - expansion'', is a powerful method which combines the
merits of perturbation theory with those of variational approaches.
The underlying idea is simple. Generically, the Lagrangian is split
into a free and an interacting part in such a way that an arbitrary
parameter $\lambda $ (or more) is artificially introduced. The
interacting part is multiplied by a factor $\delta $ which serves as
expansion parameter and is put equal to one at the end. The exact
solution should be independent of the parameter $\lambda $ while any
approximate solution will depend on it. The idea, often called
``Stevenson's principle of minimal sensitivity'' \cite{stev81}, is
that the approximate solution should depend as little upon the
parameter as possible. This means that $\lambda $ should be chosen
such that the quantity to be calculated has an extremum. In this way
the result becomes non perturbative because $\lambda $ becomes a non
linear function of the coupling constant. In every order of
perturbation theory the parameter has to be calculated again and
usually goes to infinity with growing order.

The method has been applied with great success to simple cases like
the zero dimensional and one dimensional (quantum mechanics)
anharmonic oscillator \cite{anham}, where rigorous proofs for the
(rapid) convergence exist. There are also interesting applications to
the calculation of the effective potential and the question of
spontaneous symmetry breaking \cite{oko} - \cite{stan}. In this
context
the method is usually called ``Gaussian effective potential'' (GEP),
or, in higher orders, ``post Gaussian effective potential'' (PGEP).
We also mention applications in lattice theory \cite{dun.mos}.

In \cite{oko} - \cite{stan} the mass parameter of the free
Lagrangian was treated as variational parameter. In our approach we
will fix it at the physical mass. Instead we scale the free part of
the Lagrangian with a factor $\zeta $. That's all! Due to the
``benevolent paradox'' \cite{stev84} of the linear $\delta $ -
expansion we don't need to calculate any Feynman graphs. We simply can
use existing calculations and modify the results. We find that this
method is not only quite simple but also very useful in order to
calculate the renormalized coupling constant in terms of the bare one
and the cutoff. All our results are already obtained in one-loop
order.

For $\Phi _4^4$ - theory, treated in sect. 2, and QED in sect. 3, one
easily finds that these theories are free, i.e. that the renormalized
coupling constant goes to zero when the cutoff goes to infinity,
irrespective of the behavior of the bare coupling. In the case of
$\Phi ^4_4$ - theory one can explicitly show that the conclusion
holds in any (even) order of perturbation theory in $\delta $. In QED
one can use a realistic cutoff and give an upper bound for the fine
structure constant. Yang-Mills theories are treated in sect. 4. Here
two important changes happen. Firstly, the change of the sign of the
$\beta $ - function is crucial. Secondly, gauge invariance enforces
special conditions on the splitting of the Lagrangian. Both together
implies that the theory can become non-trivial. In all cases
generalization to higher orders is straightforward in principle.

Our methods are not rigorous but they give transparent analytical
expressions in a simple way and lead to an understanding of the
relevant features of the various theories. \newpage

\setcounter{equation}{0}\addtocounter{saveeqn}{1}%

\section{$\Phi ^4_4$ - Theory}

We split the Lagrangian

\begin{equation} {\cal L}=\frac{1}{2}\partial_\mu \Phi \partial^\mu
\Phi -\frac{m_0^2}{2}\Phi ^2-\frac{g_0}{4!}\Phi ^4 \end{equation}
by introducing an artificial parameter $\zeta $ which scales the free
Lagrangian. The expansion parameter is called $\delta $ as usual. The
parameter $M$ will not be treated as a variational parameter but will
be fixed at the physical mass. This is very convenient and will lead
to the results in a simple way.

\alpheqn
\begin{equation} {\cal L} = {\cal L}_0 + \delta {\cal L}_I
\end{equation}
with
\begin{equation} {\cal L}_0=\frac{1}{\zeta }[\frac{1}{2}\partial_
\mu \Phi \partial^\mu \Phi-\frac{M^2}{2}\Phi ^2],  \end{equation}
\begin{equation} {\cal L_I}=(1-1/\zeta )[\frac{1}{2}\partial_\mu \Phi
\partial^\mu \Phi -\frac{M^2}{2}\Phi ^2]-\frac{m_0^2-M^2 }
{2}\Phi ^2 - \frac{g_0}{4!}\Phi ^4. \end{equation}
\reseteqn
For $\delta =1$ the original Lagrangian is recovered.

The Feynman rules are directly read off from (2.2). The essential
modification is that the propagator acquires a factor $\zeta $ and
that we obtain insertions containing the free Lagrangian. They will be
denoted by a thick dot in order to distinguish them from the mass
insertions which, as usual, are denoted by a cross:

\alpheqn\begin{description}
\item[Propagator:] \begin{equation} \frac{i\zeta }{p^2-M^2+i\epsilon }
\end{equation}
\item[Free Lagrangian insertion:] \begin{equation} i\delta (1-1/\zeta
)(p^2-M^2)  \end{equation}
\item[Mass insertion:]\begin{equation} -i\delta (m_0^2-M^2)
\end{equation}
\item[Vertex:] \begin{equation}  -i\delta g_0.  \end{equation}
\end{description}\reseteqn

It is easy to see how the insertions of the free Lagrangian work. If
we combine an insertion with {\em one} adjacent propagator, the
$p^2-M^2$ cancels and one is left with a factor $\delta (1-\zeta) $.
Summing up the geometrical series consisting of the bare propagator
together with $1,2,\cdots ,n$ insertions and putting $\delta =1$
gives

\begin{equation} \frac{i\zeta }{p^2-M^2+i\epsilon} \sum_{\nu =0}^n
(1-\zeta )^\nu  =\frac{i}{p^2-M^2+i\epsilon }[1-(1-\zeta )^{n+1}] .
\end{equation}
Clearly the extremum is at the ``natural'' value $\zeta =1$. For
$0<\zeta <2$ the series converges and the limit is independent of
$\zeta $ as it should be. In a theory with interactions there are, of
course, additional contributions which will shift the extremum away
from 1.

Calculations with the Lagrangian (2.2) are easily performed by just
modifying the expressions of usual perturbation theory. Although our
essential results are already obtained at the one loop level we
present the general procedure. Let $\Gamma_{2N}(p_i)$ be the connected
one particle irreducible Green function with $2N$ external legs, the
propagators for the external legs are not included. In usual
perturbation theory with respect to the bare coupling constant $g_0$,
one has an expansion of the form

\begin{equation} \Gamma_{2N}(p_i)=\sum_{V =N-1}^\infty g_0^V
\Gamma_{2N}^{(V)}(p_i).  \end{equation}
In our approach the Green function will depend on $\zeta $ and $\delta
$ in any finite order, therefore we denote it by $\Gamma_{2N}
(p_i,\zeta ,\delta )$. If we expand it with respect to $\delta $,
every internal line gets an extra factor $\zeta $. There are $I=2V -N$
internal lines in the graphs with $V$ vertices which contribute to
$\Gamma_{2N}^{(V)}(p_i)$. Furthermore, we have to take into account
the insertions of the free Lagrangian which give a factor $\delta
(1-\zeta )$ compared to the graph without the insertion. The number of
possibilities to place $J$ insertions on $I$ internal lines (compare
the well known analogous problem of Bose statistics to put $J$
indistinguishable particles into $I$ boxes) is

\begin{equation} {I+J-1 \choose J} = {2V -N+J-1 \choose J}.
\end{equation}
For the expansion of $\Gamma_{2N}(p_i,\zeta ,\delta )$ with respect to
$\delta $ up to order $n$ we therefore obtain, if we substitute $V=\nu
-J$

\begin{equation} \Gamma_{2N}(p_i,\zeta ,\delta ) = \sum_{\nu =N-1}^n
\delta ^\nu \sum_{J=0}^{J_{max}} {2\nu -N-1-J \choose J} \zeta ^{2\nu
-N-2J} (1-\zeta )^J g_0^{\nu-J} \Gamma_{2N}^{(\nu-J )}(p_i),
\end{equation}
with $J_{max}$ = Max$\{\nu -1-[N/2],0\}$. Consider next the two point
function $G_2(p^2,\zeta ,\delta )$ of all one particle irreducible
contributions to the propagator, which will be needed to calculate the
wave function renormalization $Z_\Phi (\zeta ,\delta )$. The graphs
which contribute to $G_2(p^2,\zeta ,\delta )$ up to order $\delta ^2$
are shown in fig. 1. In general one has the expansion \alpheqn

\begin{equation} G_2(p^2,\zeta ,\delta ) = \frac{i\zeta
}{p^2-M^2+i\epsilon } \left\{1 +\delta (1-\zeta )+\frac{\zeta \Sigma
(p^2,\zeta ,\delta )}{p^2-M^2+i\epsilon } \right\} \end{equation}
with
\begin{equation} \Sigma (p^2,\zeta ,\delta ) \equiv \Gamma_2(p^2,\zeta
,\delta ) = \sum_{\nu=1}^n \delta ^\nu \sum_{J=0}^{J_{max}} {2\nu -2-J
\choose J} \zeta ^{2\nu -1-2J} (1-\zeta )^J g_0^{\nu-J} \Gamma_2
^{(\nu-J )}(p^2). \end{equation}
\reseteqn
The propagator $D(p^2,\zeta ,\delta )$ is obtained by summing up the
geometrical series of all one particle irreducible contribution
contained in $G(p^2,\zeta ,\delta )$. Putting

\begin{equation} \Sigma (p^2,\zeta ,\delta )=(p^2-M^2)\Sigma '
(M^2,\zeta ,\delta )+\sigma (p^2,\zeta ,\delta ) \end{equation}
one finds

\begin{eqnarray} D(p^2,\zeta ,\delta ) & = & \frac{i\zeta }
{(p^2-M^2+i\epsilon ) [1-\delta (1-\zeta )-\zeta \Sigma '(M^2,\zeta
,\delta )]-\zeta \sigma (p^2,\zeta ,\delta )} \nonumber\\ & = &
\frac{iZ_\Phi (\zeta ,\delta )}{p^2-M^2+i\epsilon
-\Sigma_{ren}(p^2,\zeta ,\delta )},\end{eqnarray}
with
\begin{equation} Z_\Phi (\zeta ,\delta )=\frac{\zeta }{1-\delta
(1-\zeta) -\zeta \Sigma '(M^2,\zeta,\delta )} \mbox{ and }
\Sigma_{ren}(p^2,\zeta ,\delta )= Z_\Phi (\zeta ,\delta )\sigma
(p^2,\zeta ,\delta ) . \end{equation}
Besides the propagator and the wave function renormalization constant
we need the vertex function $\Gamma(p_i,\zeta ,\delta )$ (normalized
such that the expansion starts with 1). This, in turn, determines the
vertex renormalization constant $Z_V(\zeta ,\delta )$ through

\begin{equation} \Gamma(p_i,\zeta ,\delta )\rightarrow 1/Z_V(\zeta
,\delta ). \end{equation}
Since we are only interested in the behavior for cutoff to infinity,
the special choice of the momenta for the renormalization prescription
of $\Gamma$ is unessential.

In ordinary perturbation theory with respect to the bare coupling
constant $g_0$, one has the formal expansions

\begin{equation} Z_\Phi = 1 + \sum_{\nu =1}^\infty g_0^\nu Z_\Phi
^{(\nu )},\quad \Gamma(p_i) = 1 + \sum_{\nu =1}^\infty g_0^\nu \Gamma
^{(\nu )}(p_i), \quad 1/Z_V = 1 + \sum_{\nu =1}^\infty g_0^\nu
\bar{\Gamma} ^{(\nu )}, \end{equation}
where $\bar{\Gamma}^{(\nu )}$ are the coefficients $\Gamma ^{(\nu
)}(p_i)$ taken at the external momenta where the renormalization
prescription is imposed.

The corresponding expansion for $Z_\Phi (\zeta ,\delta )$ with respect
to $\delta $ is obtained from (2.11), (2.9), (2.8). The quadratically
divergent tadpole in fig. 1 is independent of the external momentum
and only contributes to the mass renormalization. If $M$ is chosen as
the physical mass, it is canceled by the mass counterterm. Therefore
we may omit all tadpole contributions here and in the following. From
the graphs in fig. 1, together with the foregoing considerations we
get

\begin{equation} Z_\Phi(\zeta ,\delta ) = \zeta \left\{ 1+\delta
[1-\zeta ] + \delta ^2 [(1-\zeta )^2+g_0^2\zeta ^4Z_\Phi
^{(2)}]+O(\delta ^3)\right\} . \end{equation}
In general, a term $\delta g_0\zeta ^2Z_\Phi^{(1)}$ would also
show up in the curly bracket of (2.14) which, however, vanishes in
$\Phi ^4$ - theory.

For the vertex functions the graphs of fig. 2 contribute. This results
in

\begin{equation} 1/Z_V(\zeta ,\delta ) = 1+ \delta g_0\zeta
^2\bar{\Gamma}^{(1)} +\delta ^2[2g_0\zeta ^2(1-\zeta
)\bar{\Gamma}^{(1)} + g_0^2\zeta ^4\bar{\Gamma }^{(2)}] +O(\delta ^3).
\end{equation}
The relation between bare and renormalized coupling constant is:

\begin{equation} g=\delta g_0 Z_\Phi^2(\zeta ,\delta ) /Z_V(\zeta
,\delta ) . \end{equation}
We need this relation only up to order $\delta ^2$ at the moment:

\begin{equation} g = \delta g_0\zeta ^2\left\{1+\delta [2(1-\zeta )
-3g_0\zeta ^2 C/2] +O(\delta ^2) \right\}. \end{equation}
The constant $C$ is defined by

\begin{equation} \bar{\Gamma}^{(1)}=-3C/2,\mbox{\quad with\quad} C
\rightarrow \frac{b}{3}\ln\frac{\Lambda ^2}{M^2} = \frac{1}{(4\pi )^2}
\ln\frac{\Lambda ^2}{M^2}\mbox{\quad for\quad } \Lambda \rightarrow
\infty, \end{equation}
and $b$ is the first coefficient in the usual expansion of the $\beta
$ - function, $\beta (g)=bg^2+\cdots$. Here and in the following we
use the results of well known one loop calculations which may be found
e. g. in \cite{iz}.

Before going on we have to clarify an important conceptional point. In
quantum mechanics or any other theory without infinities one would
prescribe $g_0$ and then calculate $g$ as a power series in $\delta $.
In the usual treatment of quantum field theory, however, one
proceeds the other way round. One fixes $g$ at its physical value and
calculates $g_0$ as a power series with coefficients that are
divergent in the limit $\Lambda \rightarrow \infty$. We stress that
the latter method is not applicable here! The reason is that the
expansion of $g$ starts with a term proportional to $\delta $.
If we invert (2.17), we obtain

\begin{equation} \delta g_0\zeta ^2 = g\{1-2\delta (1-\zeta ) + 3gC/2
+ \mbox{ rest }\}.  \end{equation}
But now the rest contains an infinity of terms of order $\delta ^0$,
because $\delta ^\nu g_0^\nu \zeta ^{2\nu } = g +\cdots$ starts with
order 1, not with order $\delta ^\nu $. Therefore it is not allowed to
truncate the series. If we do it nevertheless, this means that we
return to naive perturbation theory in $g$. In the latter case one
easily finds that the renormalized vertex function $\Gamma_{ren}(p_i)
= Z_V\Gamma(p_i)$ as well as other renormalized quantities like the
self energy, become independent of $\zeta $. This also happens in
higher orders. The renormalization procedure is powerful enough to
remove our manipulations with the Lagrangian!

We will see now, however, that the linear $\delta $-expansion in our
special formulation is extremely useful and simple in order to obtain
information about the renormalized coupling constant in terms of the
bare coupling and the cutoff. We return to (2.17), truncate after the
order $\delta ^2$, and put $\delta =1$, thus ending up with

\begin{equation} g=g_0 \{3\zeta ^2-2\zeta ^3-3g_0C\zeta ^4/2\}.
\end{equation}
The equation for the extremum in $\zeta $ (dropping the unacceptable
solution $\zeta =0$) reads

\begin{equation}  1-\zeta =g_0C\zeta ^2.  \end{equation}
Eliminating $g_0C$ from this, (2.20) becomes

\begin{equation} g=\frac{1}{2C}(1-\zeta )(3-\zeta ) . \end{equation}
Let us first assume, as usual, that $g_0>0$. Then (2.21) has just one
positive solution for $\zeta $. The solution lies between 0 and 1, in
this interval $(1-\zeta )(3-\zeta )\leq 3$ and thus

\begin{equation} g\leq\frac{3}{2C}\rightarrow \frac{3(4\pi
)^2}{2\ln(\Lambda ^2/M^2)}. \end{equation} Therefore, in the limit
$\Lambda \rightarrow \infty$ the renormalized coupling constant $g$
will necessarily converge to zero, irrespective whether $g_0$ becomes
constant, goes to zero, or diverges.

In the literature there are suggestions for possible non-trivial and
stable $\Phi ^4_4$ - theories, the ``precarious''
theory \cite{stevz}, \cite{stev85}, \cite{ste.all}, \cite{stan.stev}
($g_0<0$ infinitesimal, $g<0$ finite) and the ``autonomous'' theory
\cite{auto}, \cite{ste.all}, \cite{stan} ($g_0>0$ infinitesimal,
$g>0$ finite, infinite wave function
renormalization). As we have just seen, an autonomous theory cannot
arise in our approach. Let us look for the possibility of a precarious
theory by putting $g_0 \equiv -\gamma <0$. In this case (2.21) has two
solutions with $\zeta >1$. A finite value of $g$ in the limit $C
\rightarrow \infty $ can now be obtained if $\gamma C \rightarrow 0$,
$\zeta \rightarrow 1/\gamma C \rightarrow \infty$. Therefore the
negative bare coupling constant must become infinitesimally small. For
$g$ one then would get $ g\rightarrow 1/2\gamma ^2C^3$ which is
positive. Depending on how fast $\gamma $ vanishes in the limit
$\Lambda \rightarrow \infty $, this may converge to 0, $\infty$, or to
a finite value. We consider this possibility, however, as
unacceptable. According to the general philosophy of the principle of
minimal sensitivity the second extremum at $\zeta \approx 1+\gamma C$
should be preferred because the second derivative is (drastically!)
smaller. (This would also, and even stronger, be the case if we would
have chosen $1/\zeta $ as variation variable.) This second solution
would lead to a negative $g\approx -\gamma \rightarrow 0$. Therefore
these exotic possibilities do not show up in our approach.

It is instructive to look at the perturbative features and the
analyticity properties contained in (2.20), (2.21). One may expand the
solution (2.21) for the extremum

\begin{equation} \zeta =(\sqrt{1+4g_0C}-1)/2g_0C =
1-g_0C+2(g_0C)^2-5(g_0C)^3+14(g_0C)^4+\cdots \end{equation}
and introduce this into (2.20) to obtain

\begin{equation} g=g_0\{1-3g_0C/2+3(g_0C)^2-7(g_0C)^3+18(g_0C)^4
+\cdots \} .  \end{equation}
The linear term in $g_0C$ coincides with that of naive perturbation
theory as it should, because we expand at the extremum.

Obviously (2.24) has a branch cut at $g_0C = -1/4$, corresponding to
$\zeta =2$, which determines the radius of convergence of (2.25). We
expect that in higher orders of the optimized $\delta $ - expansion
the branch cut approaches zero as it happens in simple models. So the
method shows how the non-analytic behavior at $g_0=0$ and the
divergence of ordinary perturbation theory for any $g_0$ arises.

It is surprising that all the previous conclusions can formally be
extended to arbitrary orders $n$ of the optimized $\delta $-expansion.
Instead of (2.17) one obtains a more complicated formula of similar
structure, which again has a factor $\delta g_0\zeta ^2$ in front.
What we need is information about the behavior of the various
contributions in the limit $\Lambda \rightarrow \infty$ and about the
presence of an extremum for positive $\zeta $.

If we expand the factors $Z_\Phi ^2(\zeta ,\delta )$ and $1/Z_V(\zeta
,\delta )$ in (2.16) with respect to $\delta $ we obtain series
involving the coefficients $Z_\Phi ^{(\nu )}$ and $\bar{\Gamma}^{(\nu
)}$ in (2.13), modified by the changes performed in the Lagrangian.
Because we have chosen $M$ as the physical mass, all quadratically
divergent tadpole contributions cancel and only the logarithmic
divergences survive in the coefficients. For $\Lambda \rightarrow
\infty$ one has

\begin{equation} Z_\Phi ^{(\nu )}\sim z_\nu   (\ln\frac{\Lambda
^2}{M^2})^{\nu -1} \mbox{\quad (for \quad } \nu \geq 2),
\end{equation}
\begin{equation} \bar{\Gamma}^{(\nu )} \sim (-1)^\nu C_\nu
(\ln\frac{\Lambda ^2}{M^2})^\nu \mbox{\quad with \quad }C_\nu
>0.\end{equation}
The important point is that we know the signs of the coefficients
$C_\nu $. These signs are easily obtained from inspecting the various
factors $\pm i$ stemming from the propagators, vertices, and rotations
to euclidean space in the integration variables, for a graph
contributing to $\bar{\Gamma}^{\nu }$. The remaining integrand is then
positive definite. A specific cutoff prescription for higher order
graphs is e.g. the replacement of the euclidean propagator by
$\exp[-(P^2 +M^2)/\Lambda^2] /(P^2+M^2)$. There is no simple statement
concerning the signs of the coefficients $z_\nu $, but fortunately
these signs are unimportant. In our approach we have the same
modifications as in (2.7): Every propagator gets an additional factor
$\zeta $, this has the consequence that $Z_\Phi ^{(\nu )}$ as well as
$\bar{\Gamma}^{(\nu )}$ always appear together with a factor $\delta
^\nu \zeta ^{2\nu }$. The insertions of the free Lagrangian give
factors of $\delta (1-\zeta )$ compared to the corresponding graph
without the insertion.

The clue for the existence of an extremum is the sign of the term with
the highest power of $\zeta $ in the expansion of $g$ in (2.16). From
the previous remarks it is clear that the highest power of $\zeta $
contains no insertions. Furthermore, the coefficients of
$\bar{\Gamma}^{(\nu )}$ have one power of $\ln(\Lambda ^2/M^2)$ more
than those of $Z_\Phi ^{(\nu )}$. To find the leading term in the
coefficient for $\Lambda \rightarrow \infty$, we therefore have to
take the lowest order of $Z_\Phi ^2(\zeta ,\delta )$ and the highest
order of $1/Z_V(\zeta ,\delta )$. Therefore the term with the highest
power of $\zeta $ in the expansion of (2.16) reads

\begin{equation} \delta g_0\zeta ^2\: C_{n-1}[-\delta g_0\zeta
^2 \ln(\Lambda ^2/M^2)]^{n-1}.  \end{equation}
It has a definite sign and is negative for $g_0>0$ and even order $n$.
This term with the highest power of $\zeta $ is also the one with the
highest power of $\ln(\Lambda ^2/M^2)$.

The term with the lowest power of $\zeta $ is, correspondingly,
obtained if we take the lowest order in $Z_\Phi ^2(\zeta ,\delta )$
and $1/Z_V(\zeta ,\delta )$, and only consider the insertions. This
gives

\begin{equation} 1/Z_V(\zeta ,\delta ) \rightarrow 1 \mbox{\quad
and\quad }Z_\Phi ^2(\zeta ,\delta ) \rightarrow \frac{\zeta
^2}{[1-\delta (1-\zeta )]^2} \rightarrow \frac{\zeta ^2}{[1-\delta
]^2} \rightarrow \zeta ^2\sum_{\nu =0}^{n-1}(\nu +1)\delta ^\nu.
\end{equation}
Obviously this factor is positive. Together with the previous result
this guarantees that for even $n$ there will always be at least
one maximum at positive $\zeta $, i.e. the principle of minimal
sensitivity is applicable.

It is now easy to see what happens in the limit $\Lambda \rightarrow
\infty$. In the case that $g_0\ln(\Lambda ^2/M^2)$ stays finite (or
goes to zero) there will be a maximum at some finite $\zeta $. In
this case $g_0$ necessarily goes to zero, therefore the extra factor
$g_0$ in front as in (2.17) will imply that $g$ vanishes for $\Lambda
\rightarrow \infty $. Let us next
assume that $g_0\ln(\Lambda ^2/M^2)$ diverges for $\Lambda \rightarrow
\infty $ . Then $\zeta $ at the maximum has to go to zero, because
otherwise the highest order term would dominate all the other ones.
Therefore the expression may be simplified considerably. In any term
we only need to consider the highest power of $\ln(\Lambda ^2/M^2)$
and the lowest power of $\zeta $. Insertions therefore only give
powers of $\delta $. We end up with an expression of the form

\begin{equation} g=\delta g_0\zeta ^2\left\{\sum_{\nu =0}^{n-1}(\nu
+1)\delta ^\nu + \sum_{\nu =1}^{n-1}\delta ^\nu \sum_{j=0}^{\nu -1}
S_{\nu ,j} [\zeta ^2g_0\ln(\Lambda ^2/M^2)]^{\nu -j}\right\}.
\end{equation}
The $S_{\nu ,j}$ are uninteresting numerical coefficients, the
important point is that we know the highest one: $S_{n-1,0}
= (-1)^{n-1}C_{n-1} <0$ for $n$ even.

The rest is trivial. Putting $\zeta ^2g_0\ln(\Lambda ^2/M^2)=x$ and
setting $\delta =1$, (2.30) becomes

\begin{equation} g=\frac{x}{\ln(\Lambda
^2/M^2)}\left\{\frac{n(n+1)}{2} + \sum_{\nu =1}^{n-1} \sum_{j=0}^{\nu
-1} S_{\nu ,j} x^{\nu -j}\right\}. \end{equation}
For even $n$ there is a maximum for some finite $x$, if this is
chosen one finds that $g\rightarrow 0$ for $\Lambda \rightarrow
\infty$, i.e. the theory is trivial.

The case of a precarious theory with $g_0<0$ can be excluded in the
same way if one chooses $n$ odd.

We cannot make a general statement whether there is an extremum for
$g_0>0$ and $n$ odd or vice versa; for $n=3$ there is none in the
limit $g_0\ln(\Lambda^2/M^2)\rightarrow \infty $. This does not
matter at all because we always may choose a convenient subset of
values of $n$. It is a well known feature of the optimized $\delta $ -
expansion, which shows up already in the completely understood toy
model of a ``zero dimensional $\Phi ^4$ - partition function''
\cite{anham}, that one has to restrict to even or odd $n$,
respectively.

Of course we don't claim that the previous considerations provide a
proof that $\Phi _4^4$ - theory is free. (See \cite{phi} for this
topic.) However, they certainly give some new and alternative insight
into the problem and add further evidence for the triviality of this
theory.
\newpage
\setcounter{equation}{0}\addtocounter{saveeqn}{1}%

\section{QED}

The situation is very similar to that in $\Phi ^4$ - theory, therefore
we can concentrate on the modifications. For simplicity we work in the
Feynman gauge. The free Lagrangian

\begin{equation} {\cal L}= \bar{\psi } ( i \dslash- m_0 ) \psi
-\frac{1}{2}\partial_{\mu }A_{\nu } \partial^{\mu }A^{\nu } - e_0
\bar{\psi }\Aslash\psi \end{equation}
is split in the following way:\alpheqn

\begin{equation} {\cal L}={\cal L}_0 + {\cal L}_I(\delta ),\quad
\mbox{with} \end{equation}
\begin{equation} {\cal L}_0 = \frac{1}{\zeta _2}\bar{\psi } ( i
\dslash- M ) \psi -\frac{1}{2\zeta _3} \partial_{\mu
}A_{\nu } \partial^{\mu }A^{\nu }, \end{equation}
\begin{eqnarray} {\cal L}_I(\delta ) & = & \delta ^2[(1-\frac{1}{\zeta
_2}) \bar{\psi }( i \dslash- M) \psi -\frac{1
}{2}(1-\frac{1}{\zeta _3 }) \partial _{\mu }A_{\nu }\partial ^{\mu}
A^{\nu } +(M -m_0)\bar{\psi }\psi]\nonumber\\
& - & \delta e_0 \bar{\psi }\Aslash\psi. \end{eqnarray} \reseteqn
Some comments are appropriate here.

We have now introduced two scaling parameters, $\zeta _2$ for the free
electron Lagrangian, and $\zeta _3$ for the free photon Lagrangian.
$M$ is chosen as the physical mass of the electron. The expansion
parameter is again $\delta $ which is put equal to 1 at the end. Only
for $\delta =1$ the Lagrangians have to coincide. This freedom was
used to choose different powers of $\delta $ in the various terms of
${\cal L}_I(\delta )$, namely $\delta $ for the electron-photon vertex
but $\delta ^2$ for the free Lagrangian insertions and the mass
insertions. The reason for doing this is that in the familiar
perturbative treatment of QED the contributions to the mass counter
term $\delta m$ as well as to the renormalization constants
$Z_1,Z_2,Z_3$ always arise in connection with loop graphs, which in
lowest order are proportional to $\alpha _0=e_0^2/4\pi $. The
Lagrangian (3.2) has the corresponding structure, it is not only
invariant under the transformation $e_0 \rightarrow - e_0,\;A_\mu
\rightarrow -A_\mu $, but also under $\delta \rightarrow
-\delta,\;A_\mu \rightarrow -A_\mu $. This guarantees that at the end
expansions go with $\alpha _0$, not with $e_0$ itself.

The Feynman rules which are derived from (3.2) read:
\alpheqn\begin{description}
\item[Electron propagator:]
\begin{equation} \frac{ i \zeta _2}{\pslash-M + i \epsilon}
\end{equation}
\item[Photon propagator:]
\begin{equation} - \frac{i  \zeta_3 g_{\mu \nu }}{k^2+ i \epsilon }
\end{equation}
\item[Electron free Lagrangian insertion:]
\begin{equation} i \delta ^2(1-1/\zeta _2)(\pslash-M )\end{equation}
\item[Photon free Lagrangian insertion:]
\begin{equation} - i \delta ^2(1-1/\zeta _3)k^2g_{\mu \nu}
\end{equation}
\item[Electron mass insertion:]
\begin{equation}  i \delta ^2(M -m_0)\end{equation}
\item[Electron photon vertex:]
\begin{equation} - i \delta e_0\gamma ^{\mu }.\end{equation}
\end{description}\reseteqn

As in the previous section, Green functions can be calculated by using
the expansion coefficients of ordinary perturbation theory. Because it
is more convenient to write the expansion in $\alpha _0$ instead of
$e_0$ we introduce the following notation. Let $2N_e$ denote the
number of external electron lines, $2N_\gamma +\sigma $ with $\sigma
=$ 0 or 1 the external photon lines, and $2V+\sigma $ the vertices.
The usual perturbation expansion of the vertex function then reads

\begin{equation} \Gamma_{2N_e,2N_\gamma +\sigma }(p_i) =e_0^\sigma
\sum_{V = N_e+N_\gamma -1}^\infty \alpha _0^V \Gamma_{2N_e,2N_\gamma
+\sigma }^{(V)}(p_i). \end{equation}
The graphs which contribute to $\Gamma_{2N_e,2N_\gamma +\sigma
}^{(V)}$ have $I_e=2V+\sigma -N_e$ internal electron lines and
$I_\gamma =V -N_\gamma $ internal photon lines. Considering the
additional factors in the propagators and the insertions in analogy to
(2.7), we obtain for the expansion up to order $2n+\sigma $

\begin{eqnarray} \lefteqn{\Gamma_{2N_e,2N_\gamma +\sigma }(p_i,\zeta
_2,\zeta _3,\delta ) = }\nonumber\\
& & (\delta e_0)^\sigma \sum_{\nu =N_e+N_\gamma -1}^n \delta ^{2\nu
}\sum_{J_e,J_\gamma =0} {2\nu +\sigma -N_e-1-J_e-2J_\gamma \choose
J_e} {\nu -N_\gamma  -1 -J_e\choose J_\gamma }\times\nonumber\\
& & \times\zeta _2^{2\nu +\sigma -N_e-2J_e-2J_\gamma } (1-\zeta
_2)^{J_e} \zeta _3^{\nu -N_\gamma -J_e-J_\gamma }(1-\zeta
_3)^{J_\gamma } \alpha _0^{\nu -J_e-J_\gamma }\Gamma_{2N_e,2N_\gamma
+\sigma }^{(\nu -J_e- J_\gamma )}(p_i). \end{eqnarray}
We keep as close to the usual notation as possible. The vertex
renormalization, electron wave function renormalization, and photon
renormalization constants are now functions of $\zeta _2,\zeta _3,
\delta $ and are denoted by $Z_k(\zeta _2,\zeta _3,\delta )$. For
$\zeta _2=\zeta _3=\delta =1$ they go over into $Z_1,\;Z_2,\;Z_3$. The
formal expansion of the latter in usual perturbation theory reads

\begin{equation} Z_k = 1 +\sum_{\nu =1}^\infty Z_k^{(\nu )}\alpha
_0^\nu \mbox{\quad for\quad }k=1,2,3. \end{equation}
The Ward identity guarantees that $Z_1^{(\nu )} = Z_2^{(\nu )}$ in any
order.

The quantities $Z_k(\zeta _2,\zeta _3,\delta )$ can be expanded into
series in $\delta $. The calculation of order $\delta ^2$ is
completely parallel to the one in the last section, therefore we just
give the result:

\begin{equation} Z_1(\zeta _2,\zeta _3,\delta ) = 1+\delta ^2\zeta
_2^2\zeta _3\alpha _0Z_1^{(1)}, \end{equation}
\begin{equation}Z_2(\zeta _2,\zeta _3,\delta )=\zeta _2[1+\delta
^2(1-\zeta _2+\zeta _2^2\zeta _3\alpha _0Z_2^{(1)})], \end{equation}
\begin{equation}Z_3(\zeta _2,\zeta _3,\delta )=\zeta _3[1+\delta
^2(1-\zeta _3+\zeta _2^2\zeta _3\alpha _0Z_3^{(1)})]. \end{equation}
Clearly $Z_1(\zeta _2,\zeta _3,\delta ) \neq Z_2(\zeta _2,\zeta _3,
\delta )$, i.e. the Ward identity does not hold in it's usual form!
The reason for this is easily traced back to the extra contribution of
the free electron Lagrangian insertion in fig. 3 which contributes to
the self energy and therefore to $Z_2(\zeta _2,\zeta _3,\delta )$, but
does not appear in the vertex function and in $Z_1(\zeta _2,\zeta _3,
\delta )$. It is, however, easy to derive the modified Ward identity
from the familiar one. The result is

\begin{equation} \frac{\zeta _2}{Z_2(\zeta _2,\zeta _3,\delta )}
+\delta ^2(1-\zeta _2) = \frac{1}{Z_1(\zeta _2,\zeta _3,\delta )}.
\end{equation}
The renormalized fine structure constant $\alpha $ is obtained from
the relation

\begin{equation} \alpha =\delta ^2\alpha _0Z_2^2(\zeta _2,\zeta
_3,\delta )Z_3(\zeta _2,\zeta _3,\delta )/Z_1^2(\zeta _2,\zeta
_3,\delta ). \end{equation}
In order  $\delta ^4$ this gives

\begin{equation} \alpha = \delta^2 \alpha _0\zeta _2^2\zeta _3\left\{
1 +\delta ^2[2(1-\zeta _2) + 1-\zeta _3 +\zeta _2^2\zeta _3\alpha
_0Z_3^{(1)} \right\}.\end{equation}
Originally, instead of $Z_3^{(1)}$ one had $Z_3^{(1)} +2Z_2^{(1)}
-2Z_1^{(1)}$, the last two terms cancel due to the Ward identity. This
cancellation also happens in higher orders. Therefore, though the Ward
identity is modified in our approach, the ``old'' Ward identity for the
coefficients, $Z_1^{(\nu )} = Z_2^{(\nu )}$, still does its job.

We have to determine the extremum of (3.12) for $\delta =1$ with
respect to the two variables $\zeta _2,\;\zeta _3$. One easily sees
that

\begin{equation} \partial \alpha /\partial \zeta _2 =2 \partial \alpha
/\partial \zeta _3 \mbox{\quad for\quad } \zeta _2=\zeta
_3.\end{equation}
Thus there is an extremum with $\zeta _2=\zeta _3$. This is in
fact a necessary condition. Therefore, from now on we shall take

\begin{equation}  \zeta _2=\zeta _3\equiv \zeta .  \end{equation}
For $\delta =1$ the relation (3.12) then simplifies to

\begin{equation} \alpha = \alpha _0 \left\{ 4\zeta ^3 -3\zeta ^4
-\alpha _0C\zeta ^6\right\}. \end{equation}
Here we have defined

\begin{equation} C=-Z_3^{(1)}\rightarrow b\ln\frac{\Lambda ^2}{M^2} =
\frac{1}{3\pi }\ln\frac{\Lambda ^2}{M^2} \mbox{\quad for\quad }\Lambda
\rightarrow \infty, \end{equation}
with $b$ again being the first coefficient in the expansion of the
$\beta $ - function, $\beta (e)=b\;e\;\alpha +\cdots.$ The structure
of (3.15) is very similar to (2.20) an can be discussed along the same
lines. The maximum is at

\begin{equation}  1-\zeta =\alpha _0C\zeta ^3/2.  \end{equation}
This equation has just one real solution $\zeta $ and the solution
lies in the interval between 0 and 1. (The possibility of a precarious
theory does not arise here from the beginning, because $\alpha _0$ is
necessarily positive). Eliminating $\alpha _0C$ one obtains

\begin{equation} \alpha =\frac{2}{C}(1-\zeta )(2-\zeta )\leq
\frac{4}{C} \mbox{\quad for\quad } 0\leq \zeta \leq 1.\end{equation}
So we find that $\alpha \rightarrow 0$ for $\Lambda \rightarrow \infty
$, irrespective of the behavior of $\alpha _0$. For the triviality of
QED see \cite{lue}.

As before, one can expand the solution of (3.17),

\begin{equation} \zeta = 1-\alpha _0C/2 +3(\alpha _0C)^2/4-3(\alpha
_0C)^3/2 + 55(\alpha _0C)^4/16 +\cdots , \end{equation}
and introduce it into (3.15):

\begin{equation} \alpha =\alpha _0\{1-\alpha _0C +3(\alpha _0C)^2/2
-11(\alpha _0C)^3/4 +91(\alpha _0C)^4/16 +\cdots\}.\end{equation}
The branching point can be found by simultaneously requiring the
vanishing of the first and second derivative of (3.15). It is located
at $\alpha_0C =-8/27,\;\zeta =3/2$.

It is tempting to look into the consequences of the inequality (3.18)
if one takes it seriously. To do this we have to extend the theory by
including all elementary charged fermions and the charged $W$ - bosons
(assuming, of course, that there exist no further elementary charged
particles). We use the same factor $\zeta $ for all particles in
generalization of (3.14). In the order in which we work, the various
contributions to the vacuum polarization simply add up. We obtain
again the relations (3.15), (3.18) but now

\begin{equation} C=-\sum_fQ_f^2Z_{3,f}^{(1)}-Z_{3,W}^{(1)}.
\end{equation}
For the fermionic contributions one has, as in (3.16)

\begin{equation} -Z_{3,f}^{(1)}\sim\frac{1}{3\pi }\ln\frac{\Lambda
^2}{M_f^2}, \end{equation}
while the $W$ - boson contributes with opposite sign (see e.g.
\cite{finst}).

\begin{equation} -Z_{W,3}^{(1)}\sim -\frac{3}{4\pi}\ln\frac{\Lambda
^2}{M_W^2}. \end{equation}
For a numerical estimate we use the physical masses for leptons and
$W$, while for the quarks (to be counted three times each for color)
we take current masses as given in \cite{gas.leu}, supplemented by
the value for the top quark. To be definite we give the values used,
though they are not essential.

\begin{equation} M_u=7.6 ,M_d=13.3
,M_s=260,M_c=1270,M_b=4250,M_t=176000 MeV  .\end{equation}
For $\Lambda$ we may either choose the unification scale $\Lambda_U =
10^{15}$ GeV, or the Planck mass, $\Lambda_P = 10^{19}$ GeV, depending
on where we expect that the theory becomes modified and a natural
cutoff is provided. We then find $C_U=46.35$ and $C_P=57.59$ which
leads to

\begin{equation} \alpha \leq 0.086\mbox{\quad or\quad } \alpha \leq
0.069, \end{equation}
respectively. Because $\alpha$ is a monotonically increasing function
of $\alpha _0$ these bounds are reached for $\alpha _0\rightarrow
\infty $. The value $\alpha \approx 1/137$ would be obtained for
$\alpha _0 \approx 1/98$ and $\alpha _0\approx 1/90$ respectively.
Clearly these considerations are highly speculative at this stage.
\newpage

\setcounter{equation}{0}\addtocounter{saveeqn}{1}%

\section{Yang - Mills Theory}

The results of the last two sections, $\Phi ^4$ - theory and QED,
clearly depended on the sign of the first non-trivial correction to
the renormalized coupling constant. This, in turn, is directly
connected to the positive sign of the $\beta $ -function for small
coupling. Yang-Mills theories are asymptotically free, the
$\beta $ -function is negative and the sign of the first non-trivial
correction changes. Naively one would expect that we again obtain an
equation like (3.15) in QED, but with the sign of the term with $C$
reversed. This would imply that there is no extremum in the limit
$\alpha _0 C\rightarrow \infty $. So the method would simply avoid to
run into the previous conclusions by refusing to produce an extremum.

Actually, what happens is much more subtle. The structure of
non-abelian gauge theories will enforce a special choice of the
Lagrangian of the $\delta $ - expansion, in order that all the usual
compensations of the theory still take place. This in turn will reduce
the power of $\zeta $ in the term with $\alpha _0C$ so that
there will be an extremum now. For a suitable behavior of $\alpha _0C$
the position of this extremum moves to infinity when $C \rightarrow
\infty$. So, a finite renormalized coupling constant arises.

The usual YM - Lagrangian (the following considerations hold for any
non-abelian gauge group, for simplicity we use the language of QCD)
has the form

\begin{equation} {\cal L} = {\cal L}_0 + g_0{\cal L}_3 + g_0^2{\cal
L}_4 +g_0{\cal L}_{Gh} + g_0{\cal L}_F + {\cal L}_{GF}. \end{equation}
Here ${\cal L}_0$ denotes the free part of ${\cal L}$, ${\cal L}_3$
and ${\cal L}_4$ denote the three-gluon and four-gluon couplings,
${\cal L}_{Gh}$ the ghost Lagrangian, ${\cal L}_F$ the fermion-gluon
coupling, and ${\cal L}_{GF}$ the gauge fixing term. The explicit
expressions are well known and there is no need to repeat them here.
We use again Feynman gauge for simplicity. Remembering the result for
QED in the last section we introduce a common scaling parameter $\zeta
$ for all fields from the beginning. An essential modification is
necessary now in order not to destroy the compensations between
various graphs of the same order in $g_0$. Consider, as illustration,
the second order contributions of fig. 4 to the gluon propagator. The
three graphs in the second line have two internal lines and would get
a factor $\zeta ^2$, the graph in the third line which contains the
four gluon vertex has only one internal gluon line and would get a
factor $\zeta $ only. This would, of course, be a disaster.

Fortunately this apparent difficulty can be simply overcome by an
appropriate choice of the Lagrangian of the $\delta $ - expansion. We
introduce three functions $w_0(\zeta ,\delta ),\;w_3(\zeta ,\delta ),$
and $w_4(\zeta ,\delta )$ to be specified later, which will be
multiplied with the free part and the three-  and four-gluon couplings
respectively. Our Lagrangian reads \alpheqn

\begin{equation} {\cal L} = \frac{1}{\zeta }{\cal L}_0 + {\cal L}_I
\mbox{ with }\end{equation}
\begin{eqnarray} {\cal L}_I & = & \delta ^2 w_0(\zeta ,\delta )
(1-\frac{1}{\zeta }) {\cal L}_0 + \delta w_3(\zeta ,\delta )g_0[{\cal
L}_3 + {\cal L}_{Gh} + {\cal L}_F]\nonumber\\
&&+ \delta ^2w_4(\zeta ,\delta ) g_0^2{\cal L}_4 + {\cal L}_{GF}.
\end{eqnarray} \reseteqn
The requirement that the original Lagrangian is recovered for $\delta
$ = 1 gives the conditions

\begin{equation}  w_k(\zeta ,1) =1,  \end{equation}
but we cannot simply put all $w_k$ identical to one for the reasons
just mentioned; we have to impose the conditions imposed by gauge
invariance. Consider the vertex function $\Gamma_N(p_i)$ with $N$
external gluon lines. In ordinary perturbation theory it has the
expansion

\begin{equation} \Gamma_N(p_i)=\sum_Vg_0^V\Gamma_N^{(V)}(p_i).
\end{equation}
The number $I$ of internal gluon lines is now {\em not} fixed by $N$
and $V$, because $\Gamma_N^{(V)}(p_i)$ contains graphs with different
numbers $V_3$ of three gluon and $V_4$ of four gluon vertices.
(We forget ghosts and fermions for the moment for simplicity).
Obviously $V=V_3 + 2V_4$. We may write

\begin{equation} \Gamma_N^{(V)}(p_i)=\sum_I\Gamma_N^{(V,I)}(p_i).
\end{equation}
The numbers of vertices can be expressed in terms of $N$ and $I$:
\begin{equation}V_3=N-2V+2I,\quad V_4=3V/2-N/2-I.    \end{equation}
The expansion for the vertex function in the $\delta $ -
expansion, using the Lagrangian (4.2) reads

\begin{equation} \Gamma_N(p_i,\zeta ,\delta )=\sum_V\sum_I
[\delta w_3g_0]^{V_3} [\delta ^2w_4g_0^2]^{V_4} \zeta
^I\Gamma_N^{(V,I)}(p_i) \sum_{J=0}^{\infty } {I+J-1\choose J}[\delta
^2w_0 (1-\zeta )]^J.\end{equation}

This should be obvious from comparison with the analogous expansions
in the previous sections. The sum over $J$ which represents the
insertions, can be performed using
$\sum_J{I+J-1\choose J}x^j=(1-x)^{-I}$, and $V_3$ and $V_4$ eliminated
from (4.6). This results in

\begin{equation} \Gamma_N(p_i\zeta ,\delta ) = \sum_V \delta ^Vg_0^V
[w_3]^{N-2V}[w_4]^{3V/2-N/2} \sum_I\left\{ \frac{\zeta w_3 ^2 } {w_4
[1-\delta ^2 w_0(1-\zeta )]}\right \}^I \Gamma_N^{(V,I)}(p_i).
\end{equation}
We must insist that the compensations between various graphs of the
same order in $g_0$ have to take place also in the $\delta $ -
expansion. This will be the case if and only if all the contributions
$\Gamma_N^{(V ,I)}$ will get a factor which depends on the order $V$
only, but not on the number $I$ of internal gluon lines. Therefore the
curly bracket in (4.8) must be equal to one. This gives a relation
between the $w_k$:

\begin{equation} \zeta w_3^2(\zeta ,\delta ) = w_4(\zeta ,\delta
) [1-\delta ^2w_0(\zeta ,\delta ) (1-\zeta )]. \end{equation}
For $\delta =1$, where the $w_k$ are equal to one, this relation is
fulfilled.

It is easily seen that the inclusion of ghosts and fermions does not
alter the previous considerations because the latter always appear in
three particle vertices.

Using (4.9), the sum over $I$ in (4.8) simplifies to
$\Gamma_N^{(V)}(p_i)$. The insertions have, of course, only apparently
disappeared, they are now taken into account through the functions
$w_3$ and $w_4$ and their behavior dictated by (4.9).

There is some freedom in solving (4.9). If we insist on a reasonable
and simple solution, however, the result becomes essentially unique.
We want all $w_k$ to be polynomials in $\delta $, in order to be sure
that we don't introduce any singularities into the Lagrangian (4.2).
In order that we can take the square root of $w_3^2(\zeta ,\delta )$
in (4.9) we have to choose $w_4(\zeta ,\delta )= [1-\delta ^2
w_0(\zeta ,\delta )(1-\zeta )]/\zeta $ where the factor $1/\zeta $ is
implied by (4.3). Finally we may simply choose $w_0(\delta ,\zeta
)=1$. So we arrive at

\begin{equation} w_0(\zeta ,\delta )=1,\quad w_3(\zeta ,\delta ) =
w_4(\zeta ,\delta ) =[1-\delta ^2(1-\zeta )]/\zeta .\end{equation}
The calculation of the renormalized coupling constant $\alpha (\mu ^2)
$ at the renormalization scale $\mu $ in order $\delta ^2$ is now
straightforward along the lines of the two preceding sections. The
only difference is the consideration of the additional factors (4.10)
in the appropriate order. It is the factor of $1/\zeta $ there which
implies essential changes in the powers of $\zeta $. The result reads

\begin{equation} \alpha (\mu ^2)=\delta ^2\alpha _0\zeta \left\{1 +
\delta ^2 [(1-\zeta ) +\zeta \alpha _0C]\right\}.\end{equation}
We now have

\begin{equation}C\rightarrow 4\pi b\ln\frac{\Lambda ^2}{\mu ^2},
 \end{equation}
with $b$ the coefficient in the expansion $\beta (g) = -bg^3+\cdots$
of the $\beta $ - function. (Strictly speaking, we should better use
dimensional regularization, $d=4-2\epsilon $, now, and later
retranslate $1/\epsilon \rightarrow \ln(\Lambda ^2/\mu ^2)$.) For
$\delta =1$ (4.11) becomes

\begin{equation}\alpha (\mu ^2)=\alpha _0\{2\zeta -\zeta ^2 +\alpha
_0C \zeta ^2 \}. \end{equation}
The extremum is at $\zeta =1/(1-\alpha _0C)$. Obviously we now have
$\zeta >1$ as long as $\alpha _0C<1$. Introducing this value for
$\zeta $ gives the simple relation

\begin{equation} \alpha (\mu ^2) =\frac{\alpha _0}{1-\alpha _0C}.
\end{equation}
Now, if for $C\rightarrow \infty $ the bare coupling $\alpha _0$ goes
to zero such that $1-\alpha _0C \rightarrow \epsilon \rightarrow 0^+$
one has $\alpha \rightarrow 1/C\epsilon $ which will be finite if
$\epsilon $ is proportional to $1/C$.

To be more explicit we have to note that $\alpha _0\sim 1/C$ must not
depend upon the arbitrary renormalization scale $\mu $. Therefore we
have to introduce a dimensional parameter $\Lambda _{QCD} $ and take
$\alpha _0 \sim 1/[4\pi b\ln(\Lambda^2/\Lambda _{QCD}^2)]$. This leads
to $\epsilon = \ln(\mu ^2/\Lambda _{QCD}^2)/\ln(\Lambda ^2/\Lambda
_{QCD}^2)$ and

\begin{equation} \alpha (\mu ^2)\rightarrow \frac{1}{C\epsilon }
\rightarrow \frac{1}{4\pi b\ln(\mu ^2/\Lambda _{QCD}^2) }
\end{equation}
as it should. In a higher-order calculation it should be possible to
relate $\Lambda _{QCD}$ to the cutoff $\Lambda $.

If there were so many fermions that the theory would no longer be
asymptotically free, $b$ would be negative and the theory would become
trivial as in the examples of the previous sections. \newpage

\setcounter{equation}{0}\addtocounter{saveeqn}{1}%

\section{Conclusions}

The ansatz of scaling the whole free Lagrangian, including the kinetic
term, appears both simple and powerful. In this paper we restricted to
one loop order and to a calculation of the renormalized coupling
constant. Generalizations to higher orders and/or to the calculation
of the whole effective potential appear straightforward in
principle, without any new conceptional problems.

A characteristic of the method is, that one has to consider the bare
constant $g_0$ as given and the renormalized coupling constant $g$ as
a function of the expansion parameter $\delta $. One cannot do it the
other way round, because then an expansion with respect to $\delta $
becomes impossible. Therefore it will need some further considerations
before the method can be applied to other problems in field  theory.

Finally, one should frankly admit that all these methods of going
beyond perturbation theory have a ``distinctly alchemical flavor'' as
phrased by Duncan and Jones \cite{anham}. The splitting of the
Lagrangian into ${\cal L}_0$ and ${\cal L}_I$ is widely ambiguous,
sometimes it may be advantageous or even mandatory to use a more
complicated dependence of ${\cal L}_I$ upon $\delta $, and finally
every quantity to be calculated (even a function, say a power or a
logarithm, of the original quantity) leads to a different position of
the extremum. Nevertheless, the inherent ambiguities of the method can
be overcome by using physical principles and simplicity arguments and
lead to results which go far beyond naive perturbation theory.

\vspace*{2cm}
\noindent
{\bf Acknowledgement :} I thank P. Haberl and M. Jamin for
reading the manuscript and for valuable suggestions.

\newpage

\setcounter{equation}{0}\addtocounter{saveeqn}{1}%

\vspace*{2cm}
\noindent
{\Large \bf Figure Captions}
\bigskip\par\noindent
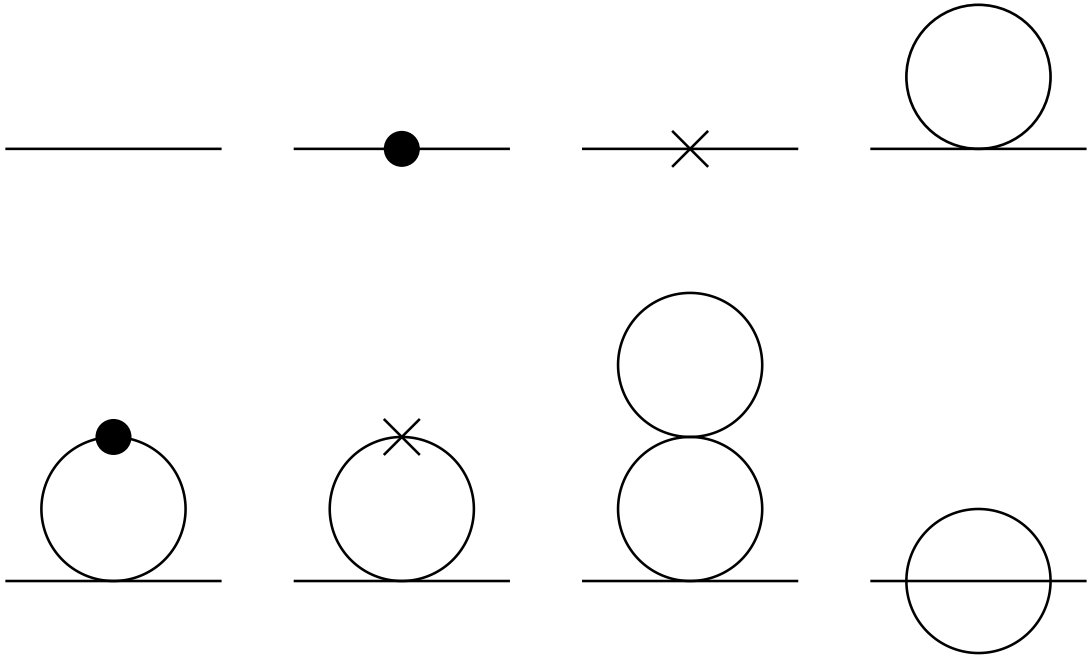
{\bf Fig. 1:} One particle irreducible contributions to the two point
function in $\Phi ^4$ - theory up to order $\delta ^2$. Here and in
the following, the thick dot denotes the insertion of the free
Lagrangian.
\par\bigskip\par\noindent
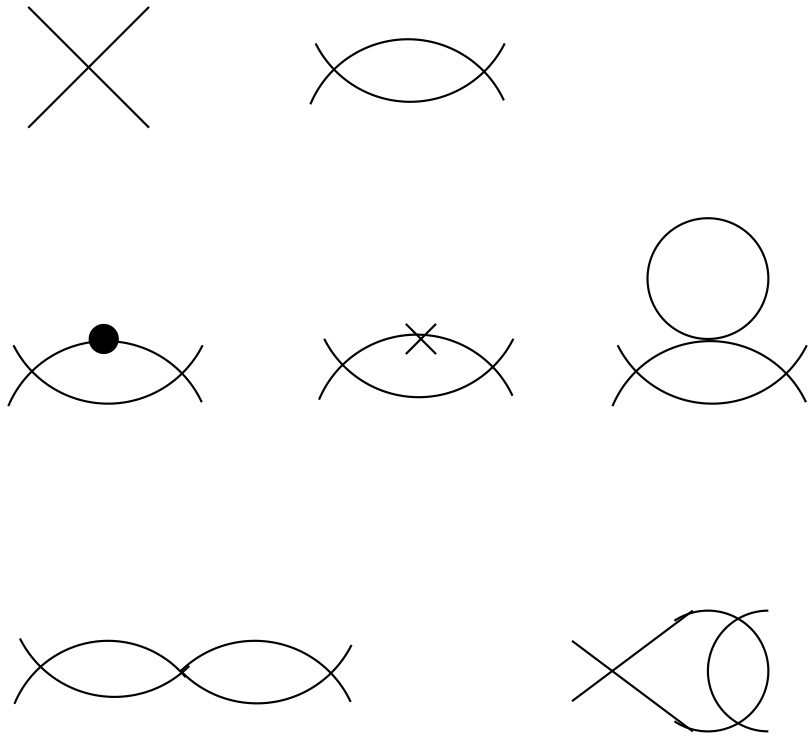
{\bf Fig. 2:} Contributions to the vertex function in $\Phi ^4$ -
theory up to order $\delta ^2$.
\par\bigskip\par\noindent
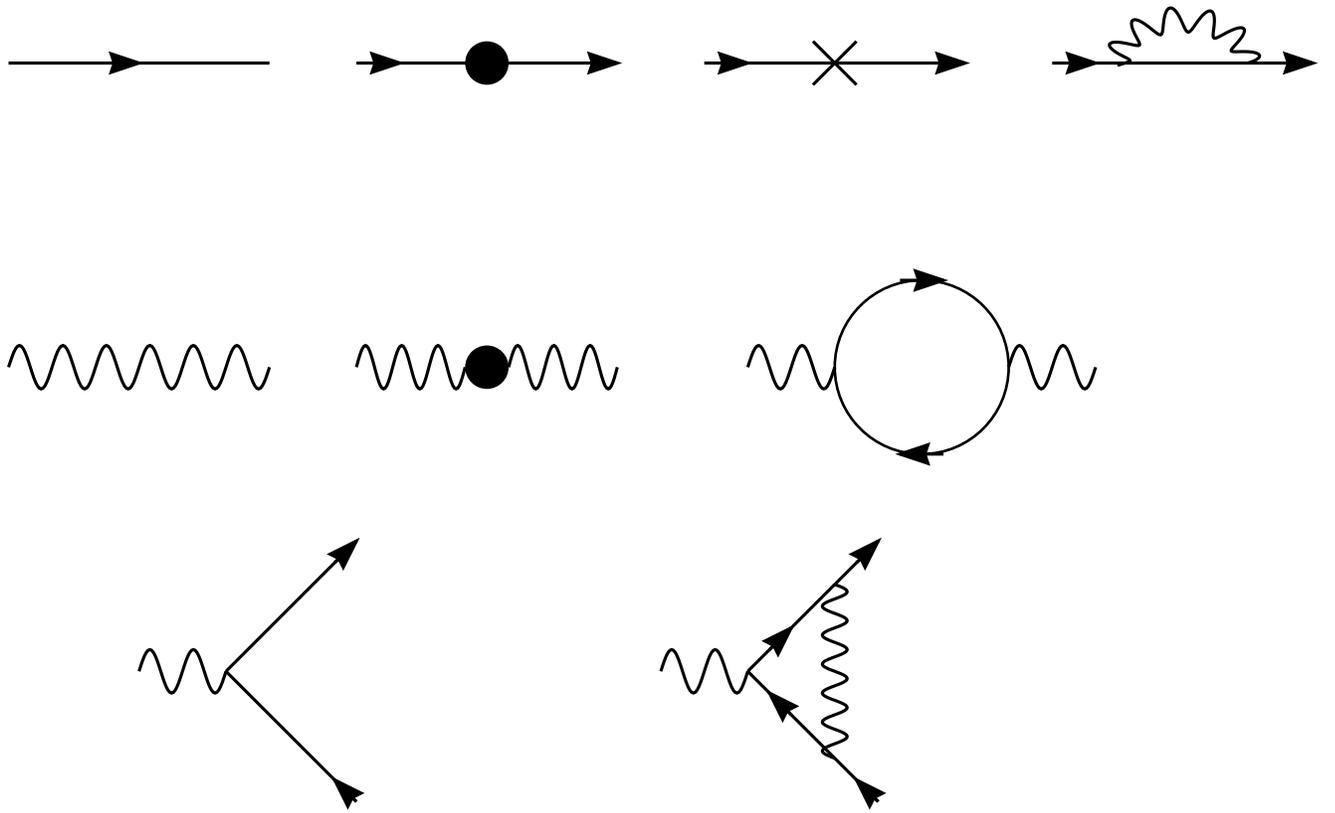
{\bf Fig. 3:} One particle irreducible contributions to the electron
two point function, photon two point function, and vertex function in
QED up to order $\delta ^2$.
\par\bigskip\par\noindent
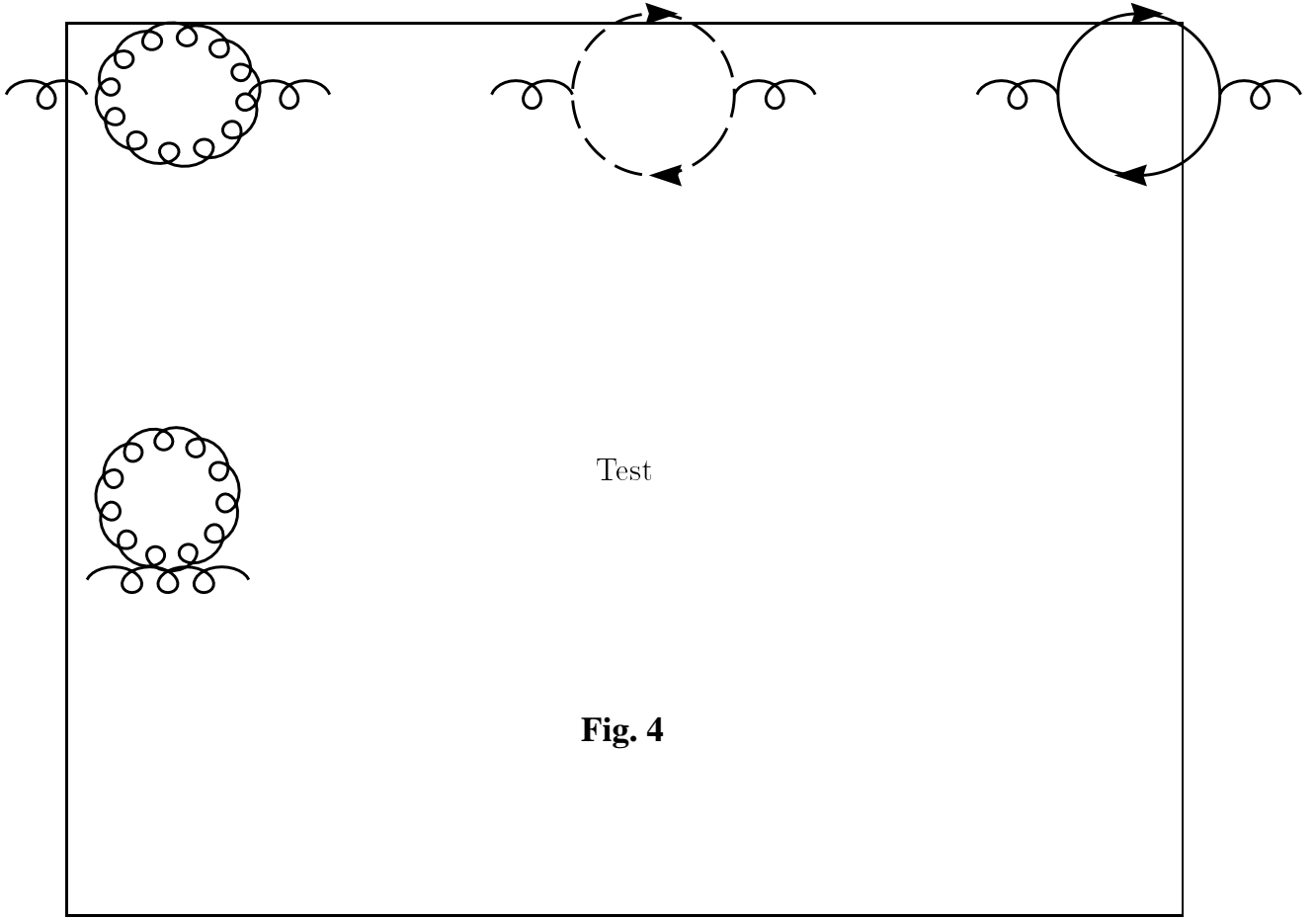
{\bf Fig. 4:} Contributions to the two point function in Yang-Mills
theory.

\end{document}